\documentclass[aps,pra,longbibliography,superscriptaddress,nofootinbib,twocolumn, floatfix]{revtex4-1}

\usepackage{times}

\usepackage{amsfonts}
\usepackage{amsmath}
\usepackage{mathtools}
\usepackage{amssymb}
\usepackage{amsthm}
\usepackage{calc}
\usepackage{physics}
\usepackage{ragged2e}

\usepackage{subfigure}
\usepackage{graphicx}
\usepackage{enumerate}
\usepackage{adjustbox}
\usepackage{appendix}
\usepackage{comment}

\usepackage[most]{tcolorbox}
\newtcolorbox{mybox}[2][]{%
  attach boxed title to top center
               = {yshift=-8pt},
  colback      = blue!5!white,
  colframe     = blue!75!black,
  halign       = flush left,
  fonttitle    = \bfseries\sffamily,
  colbacktitle = blue!85!black,
  title        = #2,#1,
  enhanced,
}

\usepackage{tikz}

\PassOptionsToPackage{pdftex,hyperfootnotes=false,pdfpagelabels}{hyperref}
\usepackage{hyperref}
\hypersetup{%
    colorlinks=true, linktocpage=true, pdfstartpage=3, pdfstartview=FitV,%
    breaklinks=true, pdfpagemode=UseNone, pageanchor=true, pdfpagemode=UseOutlines,%
    plainpages=false, bookmarksnumbered, bookmarksopen=true, bookmarksopenlevel=1,%
    hypertexnames=true, pdfhighlight=/O,
    urlcolor=blue, linkcolor=orange, citecolor=green!60!black, 
}

\makeatletter 
\hypersetup{
	pdftitle = {Quantum certification and benchmarking},
	pdfauthor = {
		Jens Eisert, 
		Dominik Hangleiter, 
		Nathan Walk,
		Ingo Roth,
		Damian Markham
		Rita Parekh, 
		Ulysse Chabaud,
		Elham Kashefi 
		},
	pdfsubject = { 
		quantum physics,
		quantum technologies,
		verification, 
		quantum supremacy,
		quantum advantage, 
		quantum computing, 
		quantum simulation,
		},
	 pdfkeywords = {quantum,
		tomography
		randomized benchmarking,
		computational complexity,
		sample complexity
	    }
	    }
\makeatother

\definecolor{jens}{rgb}{0,.8,.5}
\definecolor{ingo}{rgb}{0,.2,.9}
\definecolor{ulysse}{rgb}{.9,.5,0}

\newcommand{\je}[1]{{\color{black} #1}}
\newcommand{\ir}[1]{{\color{black} #1}}
\newcommand{\ulysse}[1]{{\color{black} #1}}

\definecolor{dominik}{RGB}{237,16,118}
\newcommand{\dom}[1]{{\color{black}#1}}

\definecolor{elham}{RGB}{255,20,100}

\definecolor{Ao}{rgb}{0.0, 0.5, 0.0}
\newcommand{\nw}[1]{{\color{black}#1}}

\begin{document}

\title{Quantum certification and benchmarking} 

\newcommand{\fu}{%
	Dahlem Center for Complex Quantum Systems,
	Freie Universit{\"a}t Berlin,
	14195 Berlin,
	Germany
}
\newcommand{\fumath}{%
	Department of Mathematics and Computer Science, Freie Universit\"at Berlin, 14195 Berlin, Germany
}
\newcommand{\hzb}{%
	Helmholtz-Zentrum Berlin f\"ur Materialien und Energie, 14109 Berlin, Germany
}
\newcommand{\pariscomp}{
	Paris  Center  for  Quantum  Computing,  CNRS  FR3640,  Paris,  France
	}
\newcommand{\sorbonne}{
Sorbonne  Universit\'e,  CNRS,  Laboratoire  d'Informatique  de  Paris  6,  F-75005  Paris,  France}
\newcommand{\edinburgh}{
	School of Informatics, University of Edinburgh, Edinburgh EH8 9AB, 
	United Kingdom
}
\author{J.\ Eisert}
\thanks{J.~E.~and D.~H.~\je{have} contributed equally.}
\affiliation{\fu}
\affiliation{\fumath}
\affiliation{\hzb}
\author{D.\ Hangleiter}
\thanks{J.~E.~and D.~H.~\je{have} contributed equally.}
\affiliation{\fu}
\author{N.\ Walk}
\author{I.\ Roth}
\affiliation{\fu}
\author{D.\ Markham}
\author{R.\ Parekh}
\author{U.\ Chabaud}
\affiliation{\pariscomp}
\affiliation{\sorbonne}
\author{E.\ Kashefi}
\affiliation{\pariscomp}
\affiliation{\sorbonne}
\affiliation{\edinburgh}

\begin{abstract}
Concomitant with the rapid development of quantum technologies, challenging demands arise concerning the certification and characterization of devices. The promises of the field can only be achieved if stringent levels of precision of components can be reached and their functioning guaranteed. This \je{review}  provides a brief overview of the known characterization methods of certification, benchmarking, and tomographic recovery of quantum states and processes, as well as their applications in quantum computing, simulation, and communication.
\end{abstract}
\maketitle

\vspace*{-.2cm}

\paragraph{Introduction.}
Recent years have seen a rapid development of quantum technologies, promising new real-world applications in communication, simulation, sensing and computation \cite{Roadmap}. Quantum internet infrastructure enables unconditionally secure transmission and manipulation of information~\cite{wehner_quantum_2018,QuantumInternet}. Highly engineered quantum devices allow for the simulation of complex quantum matter
\cite{CiracZollerSimulation}. While noisy intermediate scale quantum devices \cite{preskill2013quantum} \je{are \ulysse{on} the
verge of outperforming classical computing
capabilities~\cite{GoogleSupremacy},}
a longer term perspective of fault tolerant quantum computers 
\je{\cite{Roads}}
aims to solve impactful problems from industry that are out of reach for classical computers. These prospects come along with enormously challenging prescriptions concerning the precision with which the components of the quantum devices function. The task of ensuring the correct functioning of a quantum device in terms of the accuracy of the output
is referred to as \emph{certification} or sometimes verification.
\je{\emph{Benchmarking} more generally assigns a reproducible performance
measure to a quantum device.}

The very tasks of certification and benchmarking are challenged by intrinsic quantum features: The involved configurations spaces have enormous dimensions, a serious burden for any characterization. 
What is more, certification comes along with an ironic twist: 
It is highly non-trivial in light of the fact that certain quantum computations are expected to exponentially outperform any attempt at classically solving the same problem. 
While a large-scale universal quantum computers are still out of reach, already today do we have access to quantum simulators, that is, special-purpose, highly controlled quantum devices aimed at simulating physical systems
\cite{CiracZollerSimulation,BlochSimulation}. And, indeed, for such systems, often, no efficient classical simulation algorithm is available. As a consequence, as quantum devices are scaled up to large system sizes, application-specific tools of certification are required that go beyond standard approaches such as re-simulating a device on a classical computer or full tomographic reconstruction. It is such specifically `quantum' certification tools that this review summarizes and puts into context.

To do so, we offer a framework in which the resource cost, the information gained as well as the assumptions made in such approaches are cast very naturally. We then turn to charting the landscape of different approaches to quantum certification within our framework, ranging from practically indispensable, economic diagnostic tools such as randomized benchmarking 
to cryptographically secure techniques such as self-testing or the verification of arbitrary quantum computations in an interactive fashion~\cite{Gheorghiu2018,fitzsimons_private_2017}. 
In doing so, we aim at painting a panoramic sketch of this landscape useful for categorizing various tools and putting them into context. 
Some of the methods we lay out are crucial 
for the development and engineering of noisy near-term devices, some will find practical applications once large-scale sophisticated devices become available. 
The main importance of yet others rests in setting the stage of the possible, highlighting extremal points of this landscape, and inviting future method development to find good compromises between desirable features of a protocol.
In this review, we therefore aim to be explicit about the resource costs and assumptions made in specific protocols.

\begin{figure}[b]
\centering
\vspace*{-.7cm}
\hspace*{-.394cm}\includegraphics[width = 1.09\linewidth]{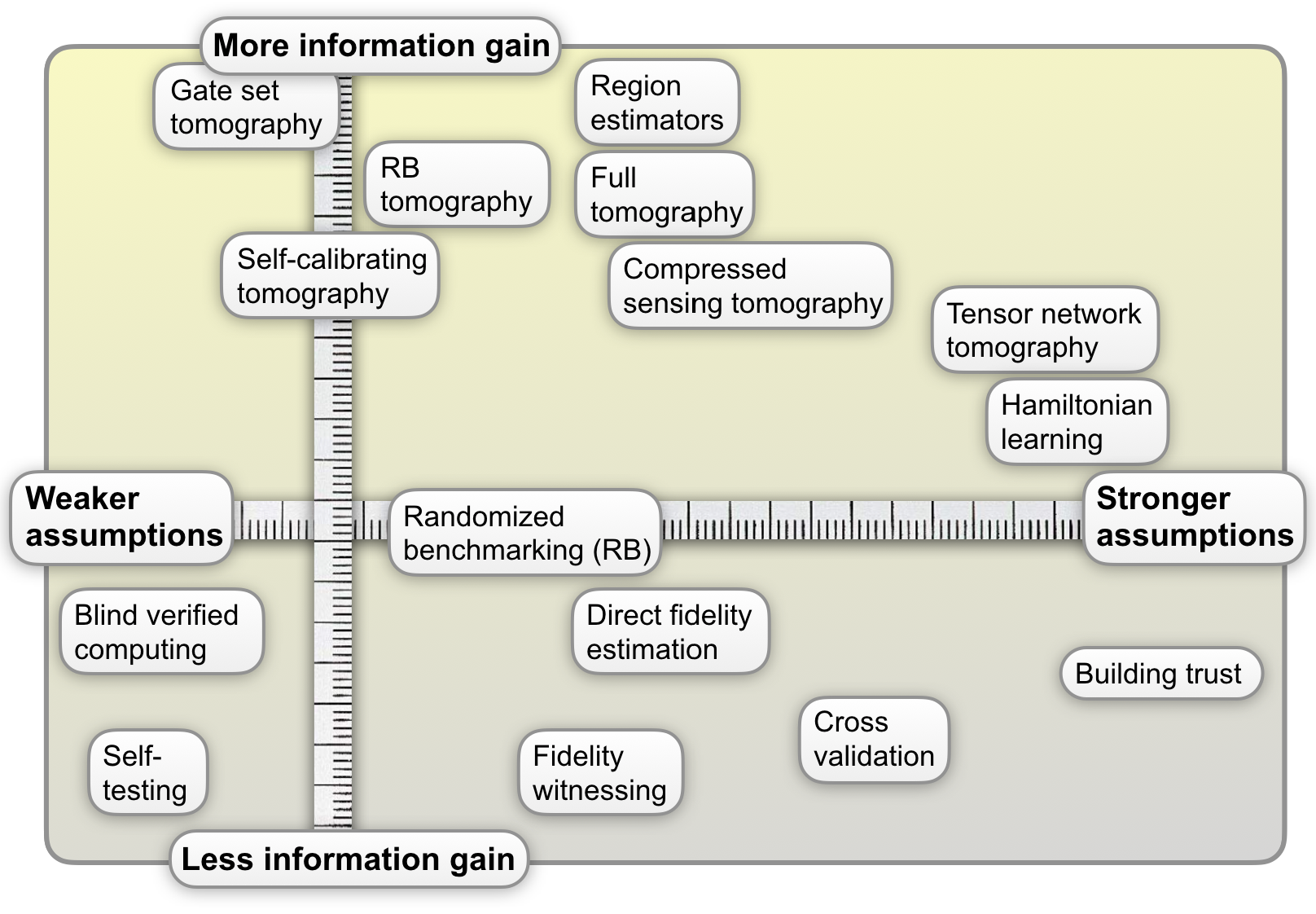}
\vspace*{-.3cm}
	\caption{Schematic of a classification scheme for some of 
	the certification protocols discussed in this article. One axis quantifies the information gain, the other the
	strength and number of assumptions required. 
	For clarity, we leave out the complexity of the protocol. An example of a protocol is discussed in \textcolor{orange}{BOX\,2}.
	}
	\label{Classification}
\end{figure}

\begin{figure*}
	
\begin{mybox}{\textcolor{orange}{BOX\,1}: Measures of quality}
\justifying{
Since a certificate should guarantee the correct functioning of a given quantum process or the correct preparation of a desired quantum state, it should ideally be phrased in terms of a measure of distance between two such objects that has an operational interpretation as their worst-case distinguishability. 
Preferably, such measures should also be \emph{composeable}, meaning that individual device or protocol certificates can be combined to certify larger, composite systems, which is especially crucial for cryptographic applications \cite{Portmann:2014uz}. Specifically, certificates for quantum states are often phrased in terms of the \emph{trace distance} $d(\sigma, \rho) = \tr[\abs{\sigma - \rho}]/2$, and for quantum channels in terms of the \emph{diamond norm} \cite{watrous_simpler_2012}, which can be conceived as a stabilized trace distance for channels. For state vectors $\ket \psi$, certificates can be easily phrased
in terms of the fidelity $F( \rho, \ket \psi\! \bra \psi ) = \bra \psi \rho \ket \psi$, which measures the overlap between $\ket \psi\!\bra \psi$ and $\rho$. 
The \emph{quality of quantum gates} is commonly expressed in terms of the 
\emph{average gate fidelity} \cite{PhysRevA.54.2614}. 
This quantity expresses the overlap of the output with the 
anticipated output of the quantum gate, in a way that is 
agnostic to the direction in Hilbert space. The fidelity for 
quantum states directly bounds the trace distance via $1- F(\sigma, \rho)^{1/2} \leq d(\sigma, \rho) \leq ({1- F(\sigma,
\rho)})^{1/2}$ and therefore certifies \emph{worst-case} performance. 
In contrast, the average gate fidelity only yields 
useful bounds \dom{that do not incur a dimension factor} for the diamond norm in certain special situations 
\cite{PhysRevLett.117.170502}
and can therefore certify performance typically only 
\emph{on average}. 
For concrete tasks at hand, other 
measures of quality may apply. 
\dom{Building upon such notions of fidelities, 
specific measures of quality have been introduced in different contexts. Examples include the cross-entropy \cite{boixo2018characterizing} or cross-entropy \je{benchmarking 
\cite{GoogleSupremacy}} aimed at verifying classical distributions or the \emph{quantum volume} \cite{QuantumVolume} aimed at capturing the quality of entire quantum circuits or gate sets.} 
In the context of quantum simulation, 
\je{as breaking up the entire scheme into
physical and conceptual building blocks is less obvious},
notions of coherence \cite{RevModPhys.89.041003}, entanglement \cite{Detection}, non-classicality \cite{NonClassical}
or purity (i.e., $\tr(\rho^2)$) are made use of. 
\je{In order to compare the quality of devices, one can define a precisely reproducible task and take
a well-defined measure of performance in this task (e.g., number of secure bits in a quantum key distribution protocol \cite{Scarani:2009kj})
as a valid figure of merit, a benchmark, in itself.}
}
\end{mybox}
\end{figure*}

\paragraph{Classifying quantum certification.}

In any task of quantum certification, the core aim is to establish the correct functioning of a quantum device. Given the enormous effort of a full tomographic characterization of quantum states and processes, in many practical applications, protocols for certification will necessarily be constrained in the available resources and at the same time governed by the advice one hopes to gain from the protocol. 
With this in mind, it is instructive to conceptualize the quantum certification problem as a protocol between the \emph{quantum device}, seen as being powerful, and its \emph{user}, who is restricted in her or his measurement devices and computational power.
One can classify schemes according to the effort and information gained, as well as in the assumptions made on the device and its user (see Fig.\ 
\ref{Classification}). 
Ultimately, even when one aims for tomographic knowledge, one may
conceive of certification as a protocol that outputs `accept' if the device functions correctly, and `reject' if it does not. 
Whether the protocol accepts or rejects is determined according to reasonable \emph{measures of quality} that are appropriate for the respective property of the device being certified \textcolor{orange}{(BOX\,1)}.

The \emph{assumptions} made on the devices and their users depend, among other aspects, on one's trust levels and the specific setting at hand. 
Typically, which assumptions are made also has an effect on the protocol's complexity in one way or the other, or even renders certification feasible in the first place.   
Conceptually speaking, there are three building blocks entering, each of which equipped with certain assumptions. 
This is, firstly, the \emph{quantum device} to be certified, a distinct and often physically separate entity. In experimental scenarios it is 
commonly reasonable to include knowledge concerning the underlying physical mechanism and potential sources of error in terms of an adequate modelling framework. However, it can also make sense to merely assume that the device is a quantum mechanical object. 
Secondly, it is useful to distinguish the \emph{quantum measurement apparata} used in the characterization, which might include state preparation and short circuits.
In an idealized setting, they may be assumed to be perfect. More practically relevant are situations in which one has a solid understanding of their functioning and characterized their efficiencies to some level. In many physical architectures, in particular in 
key platforms for quantum simulation, one 
can perform certain quantum measurements very accurately, 
but is severely limited in the \emph{type of measurements} 
that can be performed.
Thirdly, and finally, there is the \emph{classical data processing} which consumes storage capacity and processing time. 
Ultimately, any characterization provides classical numbers.
The \emph{device-independent setting} makes no assumptions at all about the measurement apparata and the device, taking into account the final data only. Such a setting is adequate, for example, if the device is a remote and untrusted quantum device that may be accessible only through the cloud.

The effort or \emph{complexity} of such a certification protocol can be divided into several distinct parts: This is 
the number of different settings or rounds in which data is obtained from the measurement device \emph{(measurement complexity)}. 
Implementing those different settings might require different \emph{quantum computational effort} as for example quantified by the length of the circuit that implements a certain measurement. 
Then, there is a minimal number of experiments and resulting samples that need to be obtained for a protocol to meaningfully succeed \emph{(sample complexity)}. 
Finally, one needs to process those samples involving classical computational effort in time and space \emph{(post-processing complexity)}. 

Often, the complexity of a protocol can be traded for the \emph{amount of information} about the device that the user can extract when running the protocol. 
Such information is crucial when it comes to designing and improving a concrete experimental setup, while it may be less important when the user's goal is merely to check the correct functioning of, say, a newly bought device, or a remote server. 
\nw{Indeed, whilst many certification techniques \je{have been} developed with specific applications in mind, the abstract criteria outlined above provide a framework to discuss the strengths, weaknesses and relevant application of these techniques in more general terms. 

For example, one can consider the relative importance of these criteria for applications on the spectrum from exploratory science, through proof-of-principle demonstrations to large scale technological implementation. 
At the exploratory end of the spectrum, information gain is at a premium as the researcher endeavors to maximise their understanding of the underlying physics. Often, such experiments are small-scale and involve well characterised measurement devices probing a relatively less understood target device. Here, the complexity of a technique will be less important and, while some assumptions regarding the measurement apparatus may be reasonable, they should be avoided as much as possible regarding the device to be certified. As we will see in the next section, this combination of desiderata would motivate the use of quantum state tomography and related techniques. 
In a proof-of-principle demonstration of a larger-scale but better controlled device the relative importance of the assumptions made and the information gained is reduced with respect to complexity involved.
%
If presented instead with a high-quality, large-scale device, efficiency will become crucial and remote users may prioritise simple-to-use certification techniques 
such as self-testing or benchmarking at the level of applications rather than hardware.

}

We now present and assess various tools for characterisation ordered according to, first, the information that may be extracted from the protocol, and, second, the assumptions made in the protocol. 
\dom{Rather than being exhaustive and technically detailed, our selection highlights distinct points within our framework with the goal to sketch a panoramic view of the landscape it gives rise to.} 
In addition to the main text, we provide a tabular overview in which we quantitatively assess exemplary certification protocols for applications in cloud computing, demonstrating a quantum advantage, and quantum simulation and computation according to our classification \textcolor{orange}{(TABLE\,\ref{thetable})}.
\dom{
We illustrate how to read this table by means of exemplary cases in \textcolor{orange}{BOX\,4}.
}

\paragraph{Certification protocols.}
In many scenarios, it is reasonable to assume that one's quantum measurements are rather well characterized and that the object of interest is either a quantum state or process that can be accessed in independently identically distributed (i.i.d.) experiments. 
These assumptions are often very natural in laboratory settings in which the quantum device can be directly accessed. 
They are therefore at the heart of many characterization protocols and shall be our starting point for now.

\begin{figure}
\begin{mybox}{\textcolor{orange}{BOX\,2}: Randomized benchmarking}
%
\justifying{\emph{Randomized benchmarking (RB)}
refers to a collection of
methods that aim at reliably estimating the magnitude of an
average error of a quantum gate set in robust fashion against \emph{state preparation and measurement (SPAM)} error. 
\je{It achieves} this goal by applying sequences of feasible
quantum gates of varying length, so that small errors
are amplified with the sequence length leading.
\je{From a pragmatic point of view, RB protocols thereby define benchmarks that can be used to compare different digital quantum devices.
In \je{important} instances, the benchmark can be related to the average gate fidelity, rendering RB protocols flexible certification tools.}
\je{To this end,} a group structure of the gate set is made use 
to achieve two goals: On the one hand, this is to 
control the theoretical prediction of error-free sequences. On the 
other hand, this allows one to analyze the error contribution after averaging using representation theory. 
\je{Originally devised for random unitary gates \cite{FirstRB, PhysRevA.80.012304, PhysRevA.75.022314}, RB is most prominently
considered for Clifford gates }
\cite{KnillBenchmarking, MagGamEmer2},
\je{and} has been extended to 
\je{other finite groups}
\cite{PhysRevA.90.030303,PhysRevA.92.060302,BeyondCliffordRB,PhysRevLett.123.060501, HelsenEtAl:2019:character, CycleBenchmarking}. 
Assumptions on having identical noise levels per gate 
 have been lessened \cite{IndependentNoise}, and RB with confidence introduced
\cite{RBFlammia,helsen_multiqubit_2019}.
RB schemes have been generalized to other measures of quality, such as relative average gate fidelities \cite{magesan2012efficient} with specific target gates, fidelities per symmetry sector \cite{PhysRevA.92.060302,PhysRevLett.123.060501}, 
the unitarity \cite{wallman2015estimating},
 \ir{
 measures for losses, leakage, addressibility 
 and 
 cross-talk \cite{GambettaEtAl:2012:simultaneousRB, WallmanEtAl:2015:LossRates,WallmanEtAl:2016:Leakage}
 }
 or 
even tomographic schemes that 
combine data from multiple RB experiments
\cite{KimmOhki,AverageGateFidelities, 2019arXiv190712976F}.
\je{In addition, RB protocols have been devised that directly work on the level of generating gate sets \cite{Hashagen,PhysRevLett.123.030503}.}
}
\end{mybox}
\end{figure}

The most powerful but at the same time most resource-intense such technique of certification is full \emph{quantum tomography} \cite{Hradil:1997tt,James:2001bb}. 
Here, the idea is to obtain knowledge of the full quantum state or process by performing sufficiently many (trusted) measurements. 
Given tomographic data, one can in particular obtain a certificate that the state lies in some region in state space. 
For many years such regions were typically constructed heuristically by first applying maximum likelihood estimation to construct a 
point estimate of the state \cite{hradil_3_2004} and then using resampling techniques to obtain error bars. 
More recently, techniques to obtain more rigorous region estimates have appeared including \emph{Bayesian credibility regions} \cite{BlumeKohout:2010cm,Ferrie:2014cs} and \emph{confidence regions} \cite{blume-kohout_robust_2012,christandl_reliable_2012,Wang:2019kp} where the former are usually smaller but depend strongly upon the Bayesian prior. Most importantly, from these tomographic reconstructions, one exactly learns the nature of deviation of the imperfect implementation to the target. 
Such data proves crucial when designing experimental setups as it yields information about the particular sources of errors present in the setup and hence functions as `actionable advice' on how to improve the setup.

\begin{figure*}
\begin{mybox}{\textcolor{orange}{BOX\,3}: Certifying a quantum advantage
or quantum computational supremacy}
\justifying{
Using a quantum computer to efficiently perform computational tasks that are provably intractable for classical computers marks a key milestone in the development of quantum technologies. 
Various sub-universal 
models of quantum computing have been proposed to demonstrate, with near-term achievable technology, 
a 
so-called \textit{quantum advantage} or
\textit{quantum computational supremacy}.
A crucial part of the demonstration of this claim with a given model is the verification of the output of the corresponding quantum device. 
But the nature of the computational task is precisely such that it cannot be reproduced classically and therefore the traditional means of verifying a computation fail. 
What is more, the proposed sub-universal quantum devices produce samples from exponentially flat probability distributions to the effect that it requires exponentially many samples to classically verify that the obtained samples are indeed distributed according to the target distribution, independently of the hardness of producing the samples~\cite{valiant_automatic_2017,hangleiter2018sample}.
The latter result severely restricts the possibilities for deriving classical verification protocols for quantum computational supremacy even under the assumption that the verifier has access to \emph{arbitrary computational power}.

To circumvent this no-go result and arrive at a sample-efficient verification protocol one may take very different routes:  
First, one might ask for less than verification of the full output distribution such as merely distinguishing against the uniform or certain efficiently sampleable distributions, which can often be done in a computationally efficient way~\cite{aaronson_bosonsampling_2013,spagnolo_efficient_2014,carolan_experimental_2014,Phillips:2019ho}. 
Allowing for exponential time in classical post-processing, one can also sample-efficiently verify coarse-grained versions of the target distribution~\cite{Bouland}, make use of certain complexity-theoretic assumptions~\cite{aaronson_complexity-theoretic_2016}, 
or assumptions on the noise in the quantum device \cite{boixo2018characterizing,Bouland,ferracin2018verifying}. 
The latter allows one to use weaker measures, like the 
\emph{cross-entropy} \cite{boixo2018characterizing,Bouland} or variants thereof such as the cross-entropy benchmarking fidelity~\cite{GoogleSupremacy}.
If one gives qualitatively more power to the user, e.g., trusted single-qubit measurements \cite{hangleiter_direct_2017,PhysRevLett.120.170502,takeuchi_verification_2018}, this even allows one to fully efficiently verify the prepared quantum state and thereby the sampled distribution.
Finally, one may use more complicated, interactive protocols which require a universal quantum device, e.g., the one presented in Ref.~\cite{mahadev_classical_2018}, which relies on the post-quantum security of a certain computational task to classically delegate a universal computation. 
Given the importance of verifying a quantum advantage, it is a pressing challenge to derive fully efficient verification protocols which involve minimal assumptions.
We expect that this will require custom-tailored techniques for the different available proposals. 
}

\end{mybox}
    \label{fig:Qadv}
\end{figure*}

However, generic quantum state and process tomography 
is excessively costly in the size of the quantum system. 
Fortunately, many quantum states and processes that are encountered in realistic experiments exhibit significant structure: 
States are often close to being pure or have approximately low rank, so that methods of \emph{compressed sensing tomography} \cite{Compressed,Kalev2015,2018arXiv180911162G} can be applied in which less resource expensive or more reliable recovery is possible based on the same type of (but 
randomly chosen) measurements compared to full tomography. Similarly, quantum processes are often close to being unitary \cite{Compressed2,Guaranteed}. 
For local Hamiltonian systems, even further structure of locality comes into play. In particular, tensor network states can provide meaningful variational sets for \emph{tensor network tomography}, which basically makes the structural assumption that there is little entanglement in the state, an assumption that is often valid for 
quantum many-body states to an extraordinarily good approximation \cite{Cramer-NatComm-2010,Wick_MPS,MPOTomography,Efficient}. 
Also, variational sets inspired by \emph{machine learning}
have been considered \cite{torlai_many-body_2018,carrasquilla_reconstructing_2019}.
In such situations, the effort of quantum state and process 
tomography 
can be significantly reduced. 
At least for intermediate-sized systems, 
such techniques are practically highly important. 

If one is only interested in certain properties of a quantum state or process one may resort to so-called \emph{learning techniques}, which scale much more favourably. For instance, one may merely be interested in
\emph{probably approximately correctly (PAC)} 
learning the expected outcomes of a certain set of 
measurements, e.g., local observables on the quantum state. 
PAC learning is possible with a measurement complexity that scales only linearly (in the number of qubits) in certain settings~\cite{aaronson_learnability_2007,rocchetto_stabiliser_2017,rocchetto_experimental_2019} but still incurs exponential computational effort. 
In another instance of learning, one might be confident that the given data is described by a certain restricted 
Hamiltonian (or Liouvillian) model whose parameters are however not known. 
\emph{Hamiltonian (or Liouvillian) 
learning techniques} solve this task and 
recover the Hamiltonian parameters from 
\je{suitable}
data
\cite{holzapfel_scalable_2015,GrenadeLearning}.

In contrast to the \je{aforementioned} tools for characterizing a 
quantum device, \emph{fidelity estimation} aims merely at 
determining the overlap of the actual quantum state or 
process implemented in a given setup with the ideal one.
While fidelity estimation yields much less information than full tomography, one saves tremendously in measurement and sample complexity. 
In fact, using \emph{importance sampling} one can estimate the fidelity of an imperfect preparation of certain pure quantum states in constant measurement complexity~\cite{FidelityEstimation}.
\dom{This protocol can be extended to optimally estimating the fidelity of quantum channels ~\cite{reich_optimal_2013}. 
}

A yet weaker notion than fidelity estimation is \emph{fidelity witnessing}. 
The idea of a fidelity witness is to cut a hyper-plane through quantum state space which separates states close in fidelity to a target state from those far away.
Efficient fidelity witnesses can often be derived in settings in which the target state satisfies some extremality property so that it lies in a low-dimensional corner of state space, such as certain multi-partite entangled states \cite{PhysRevLett.120.170502}, Gaussian bosonic states \cite{Leandro} or ground states of local Hamiltonians \cite{hangleiter_direct_2017}. 

A still weaker approach merely aims at verifying or estimating 
the presence of certain key properties, such as entanglement from realistic measurements, to, say, observe entanglement propagation \cite{BlattEntanglementPropagation}.
Here again, notions of (quantitative) witnesses that provide bounds to entanglement measures play an important role \cite{quant-ph/0607167,Audenaert06,Guehne}.
Such witnesses can be measured by exploiting randomness~\cite{RandomEntanglement}.

\begin{figure*}
\begin{mybox}{\textcolor{orange}{BOX\,4}: A guide to \textcolor{orange}{TABLE\,1}}
\justifying{
\dom{
In \textcolor{orange}{TABLE\,1}, 
we are comparing a wide range of protocols some of which are structurally distinct.
We have settled on certain criteria described above to meaningfully compare a variety of techniques in a unified language. 
But of course more fine-grained distinctions are necessary to exhaustively describe all the protocols.
For example, in delegated computing protocols natural figures of merit include the number of communication rounds and transmitted bits as well as overhead in terms of qubit number. 
However, such quantities do not appear in state tomography protocols. 
As the number of qubits and rounds affects the number of single-qubit measurements that need to be performed by the server in a blind computing protocol our criteria capture the effort required to perform the protocol. 
Let us give \ulysse{provide two} examples. 

\emph{Example 1} (Blind computing via trap qubits~\cite{fitzsimons_unconditionally_2017}):  Here, the client prepares and sequentially transmits to the server a product quantum state of $N \in O(n D \log(1/\epsilon)$ many qubits in order to delegate and verify a depth-$D$ quantum computation on $n$ qubits with trace-distance error $\epsilon$. 
The server entangles the qubits in a graph state, measures all of them, and sends the outcomes to the client for post-processing who simply compares certain outcomes.
The number of distinct single-qubit measurements is therefore given by one choice of settings for a single $N$-qubit measurement, while the number of samples is $1$ as the protocol is single-shot. 
To obtain the certificate, $O(N)$ many of the single-qubit measurement outcomes need to be compared and hence the post-processing is linear in that number. 
}

\ir{
\emph{Example 2} (Low-rank state tomography with $2$-designs~\cite{Guta}):
Here, one repeatedly measures a \je{positive operator valued measure}
(POVM) that constitutes a complex \je{projective} $2$-design on $O(2^nr^2/\epsilon^2)$ i.i.d.\ copies of an $n$-qubit rank-$r$ quantum state. 
Such a POVM consists of at least $O(4^n)$ elements and, hence, requires an exponential number of observables. The sampling complexity is of order of the degrees of freedom of a rank-$r$ state up to an additional factor of $r$. 
From the frequency estimates one calculates a linear least-square estimator and subsequently projects the result onto quantum states which requires a time complexity of $O(2^{3n})$ on a classical computer. 
A trace-norm ball around the obtained estimate is a confidence region depending only on the estimate's rank.
}
}
\end{mybox}
\end{figure*}
In case one has a good understanding of the physical mechanisms governing the device, 
it is often useful to \emph{build trust} in the quantum device.
This approach is particularly prominent in the context of quantum simulation:
Here, the idea is to certify a quantum device by validating its correct 
functioning in certain classically simulable regimes through 
comparison to classical simulations \cite{trotzky_probing_2012,ValidatingSimulator,schreiber_observation_2015,braun_emergence_2015}. In some instances, stronger statements can be made when
invoking notions of \emph{self-validation}~\cite{Selfverification} or \emph{cross-platform verification}~\cite{elben_cross-platform_2019}. It is also common to certify the components of a device, for example, individual gates, and extend the trust obtained in this way to the full device, making the assumption that all sources of errors are already present for the individual components. 
In such approaches, it is assumed that no additional sources of errors arise when moving out of the strictly certifiable regime again.

An important drawback of most schemes discussed so far, however, is that they assume i.i.d.\ state preparations.
This limitation can be overcome using \emph{quantum de Finetti arguments} to obtain non-i.i.d.\ tomographic regions \cite{christandl_reliable_2012,Wang:2019kp} and distance certificates~\cite{takeuchi_verification_2018}, the use of which has been optimized in various works for the case of \emph{graph states} \cite{Markham,takeuchi_resource-efficient_2018} as well as for \emph{continuous variable 
states} \cite{Chabaud19}. 

More severely still, in the standard setting a high level of trust in the measurement devices is required giving rise to a vicious cycle: 
After all, to calibrate the measurement devices in the first place, one requires quantum probe states which are well characterized, a task that requires well calibrated measurement devices.
This raises the question whether one can simultaneously learn about the quantum device and the quantum measurement apparatus in a self-consistent or semi-device-dependent way. 
The rather extreme and resource-intense solution to this problem is \emph{gate set tomography} 
which instead of focusing on a single quantum channel or state, characterizes an entire set of quantum gates, the state preparation and the measurement 
self-consistently from different gate sequences~\cite{merkel_self-consistent_2013,blume-kohout_robust_2013,blume-kohout_demonstration_2017}. 
Other solutions have been 
demonstrated in optics settings where one can perform state tomography 
in a \emph{self-calibrating way}~\cite{branczyk_self-calibrating_2012,mogilevtsev_self-calibration_2012}.
Such schemes at times even come with error bars~\cite{sim_proper_2019}. 
One can also exploit well characterized reference states such as coherent states~\cite{motka_efficient_2014} as a lever to perform uncalibrated tomography~\cite{rehacek_operational_2010}. In another vein, one can 
mitigate uncertainty in the model that generated the data by using 
model averaging techniques~\cite{ferrie_quantum_2014}. 
A particularly important example of fidelity-estimation protocols for quantum processes that break this vicious cycle \je{have been 
proposed in the context of} \emph{randomized benchmarking}~\cite{FirstRB,PhysRevA.80.012304, PhysRevA.75.022314,MagGamEmer2} 
(see \textcolor{orange}{(BOX\,2)}).

\ir{If a quantum device already allows for a level of abstraction that is close to an \je{envisioned} application, one can use 
more specific or even 
\ulysse{ad hoc} performance measures in simple tasks as a benchmark.  For example, quantum key distribution is usually certified entirely at the application level through the number of secure distributed bits \cite{Scarani:2009kj}. One can also run small instances of quantum algorithms on prototypes of digital quantum computers \cite{DebnathEtAl:2016:Programs, LinkeEtAl:2017:ProgramComparison}.
}

One could imagine, however, applications where even mild assumptions cannot be guaranteed.  In such a scenario, one could utilize a range of cryptographic tool-kits to ensure that the above assumptions are indeed enforced. 
One prominent example of such setting is where one has to work in a black-box setting, i.e., 
with no assumptions made about the underlying devices. Remarkably, the non-local correlations demonstrated by quantum mechanics allow for certain entangled states and non-commuting measurements to be certified in this setting (up to local isometries) solely via the observed statistics \cite{mayersyao04}. 
This procedure of \emph{self-testing} is typically achieved through the violation of a Bell inequality, with the paradigmatic example being the maximal violation of the CHSH inequality which self-tests non-commuting Pauli measurements made upon a maximally entangled pair of qubits. A substantial body of literature has extended these results in many directions, including generalisations to multi-partite entangled states \cite{mckague_self-testing_2011}, gates and instruments \cite{Sekatski:2018eo}, and approximate collections of maximally entangled qubit pairs \cite{Reichardt-Nature-2013} which have been made increasingly robust \cite{natarajan_low-degree_2018}
(a recent and comprehensive review of self-testing can be found in
Ref.~\cite{Supic:2019vd}).

In the context of computation, the key idea of device-independent schemes is to \emph{hide} the delegated computation from a remote black box server in such a way that the powerful server cannot retrieve any information about the computation without leaking to the client that a deviation has occurred. The use of the \emph{quantum twirling lemma} or similar technique allows one to simplify the analysis under a general deviation (with no assumptions) to a simple (i.i.d.) case leading to efficient \emph{verified blind quantum computation} schemes \cite{fitzsimons_unconditionally_2017,Reichardt-Nature-2013,coladangelo_verifier---leash:_2017}. Guarantees of correctness have been achieved in this manner in various scenarios, giving different degrees of control to the user. 
These powerful verification schemes, while removing many trust assumptions and providing efficient protocols, remain only applicable to a remote verifier with limited quantum capacity, such as single qubit gates \cite{fitzsimons_unconditionally_2017}, or access to entangled servers \cite{Reichardt-Nature-2013,coladangelo_verifier---leash:_2017}. These last obstacles have recently been overcome by utilizing yet another cryptographic toolkit, this time from the classical domain. The usage of post-quantum secure collision-resistant hash functions has enabled a fully classical client to hide and verify the remote quantum computation \cite{mahadev_classical_2018,gheorghiu_computationally_2019}. However, 
these new schemes come with a significant overhead that can be reduced to some extent~\cite{alagic_two-message_2019}, and they are no longer 
fully unconditionally secure, as they are based on a computational assumption, that is, the existence 
of classical problems that are computationally hard to solve even for a quantum 
computer~\cite{regev2010learning}.
We provide a case study of how different certification methods can be applied in the context of verifying a quantum computational advantage \textcolor{orange}{(BOX\,3)}.

\paragraph{Outlook.} In this \dom{review}, we have provided an overview of
methods for certifying and benchmarking quantum devices as they 
are increasingly becoming of key importance
in the emerging quantum technologies
(for detailed, up-to-date information, see 
\textcolor{orange}{(BOX\,5)}). 
\dom{
We hope that our framework will prove to be a useful means of \je{assessing} 
and putting into a holistic context methods to be developed in the future. 
Indeed, achieving good compromises between resource cost, obtainable information and assumptions made in a protocol may well 
be a make-or-break topic for quantum
technologies.
In this mindset,} \je{this review} -- 
\je{and \ulysse{the} quantitative framework provided here}
-- 
is also 
meant to be an invitation and guideline for future method development to a growing 
field of research that combines sophisticated
mathematical reasoning with a data-driven
experimental mindset.

\paragraph{Acknowledgements.}
We gratefully acknowledge discussions with D.~Gross
\je{and J.~Helsen, in addition to many other members of the scientific
community.}
J.~E.~acknowledges funding from
the DFG (CRC 183 project B01, EI 519/9-1, EI 519/14-1, EI 519/15-1, MATH+, EF1-7), the BMWF (Q.Link.X), 
the BMWi (PlanQK), and the Templeton Foundation. N.~W.\ 
acknowledges funding support from the European Unions Horizon 2020 research and innovation programme under the Marie Sklodowska-Curie grant agreement No.~750905.
This work has also received funding from the European 
Unions Horizon 2020 research and innovation 
programme under grant agreement No.~817482 (PASQuanS). E.~K.\
and D.~M.\
\je{acknowledge} funding from the ANR project ANR-13-BS04-0014 COMB,
E.~K.\ by the 
EPSRC (EP/N003829/1).

\begin{figure}[h]
\begin{mybox}{\textcolor{orange}{BOX\,5}: Online certification library}
\justifying{Some of the present authors have 
curated an online library of certification protocols on the
Quantum Protocol Zoo, hosted at
\href{https://wiki.veriqloud.fr}{wiki.veriqloud.fr} under certification library.
The
aim of this repository is to provide a compact and precise review of
the existing certification techniques and the 
corresponding protocols beyond the scope of this work.
As of today, this library consists of a few concrete
protocols in a specified format, classified in different techniques,
where the technique page also includes a brief description, properties
and the references. For every protocol, we provide its detailed
outline, the assumptions considered, resources and requirements, a
mathematical description of the procedure and properties including
sample and measurement complexity. Certification plays an important
role in the development of quantum devices and we hope this library
will help the community classify the certification techniques and also
keep updating them with the progress in this field.}
\end{mybox}
    \label{fig:Library}
\end{figure}

\begin{turnpage}

\begin{table*}
	\begin{tabular}{p{.14\textheight} | p{.2\textheight} | p{.09\textheight} | p{.1\textheight} |p{.08\textheight} |p{.08\textheight} |p{.09\textheight} | p{.17\textheight} }
		\toprule
		\emph{Application} &  \emph{
		Verification protocol} & \emph{Information} & \multicolumn{4}{l|}{\emph{Complexity}} & \emph{Assumptions}\\ 
		& & & Measurements & Feasibility &  Samples & Post-processing & \\
		\toprule
      \multicolumn{7}{l}{\qquad \textbf{Quantum networks and cloud computing}} \tabularnewline \hline
      Cloud quantum computing &  Blind computing via trap qubits \cite{fitzsimons_unconditionally_2017} & Trace distance & $1 \times $ \quad $O(nD \log(1/\epsilon))$ &single-qubit preparation& $1$ & $O(nD \log(1/\epsilon))$& Client prepares and sends single qubits, receives classical data\\
      & Blind computing via Bell test \cite{coladangelo_verifier---leash:_2017}& Trace distance &$O(D)\times$ \quad $  \tilde O(nD \log( 1/\epsilon))$ & classical &$1$ & $O(nD\log(nD)$ \quad \quad$\cdot \log(1/\epsilon))$ & Non-communicating and spatially separated servers \\
      & Classically verified quantum computing~\cite{mahadev_classical_2018}   &Trace distance &  $2 \times $ \quad $ O(n^3 D^2 \log(1/\epsilon)  $ & classical & $1 $ & $O(n^3 D^2 \cdot $ \quad \quad  $ \cdot  \log(1/ \epsilon))$ &\emph{Learning With Errors} problem is hard for quantum computers\\
Secret sharing & Graph state schemes \cite{Markham}  & Trace distance& $O(1/\epsilon) \times n$ &Pauli obs.&  $1$ & $O(n/\epsilon)$ & i.i.d., trusted single-qubit measurements for honest players \\\hline
	\multicolumn{7}{l}{\qquad \textbf{Quantum advantages}} \\\hline
	 Boson sampling & State discrimination \cite{aaronson_bosonsampling_2013,carolan_experimental_2014,spagnolo_efficient_2014} & Discriminate from uniform   & $1 \times n$ & classical & $O(1)$ & $O(n)$ & Characterized random unitary\\
	 Random circuit sampling  & HOG verification \cite{aaronson_complexity-theoretic_2016} & Classically hard task achieved  & $1 \times n $ & classical& $O(1)$ & $O(\exp(n))$ & Generating larger-than-median outputs is classically hard. \\
	 & Cross-entropy verification \cite{boixo2018characterizing,Bouland} & TV distance  & $1 \times n$ & classical & $O(n^2/\epsilon^2)$ & $O(\exp(n))$ &  $H(p_{\mathrm{device}}) \geq H(p_{\mathrm{ideal}})$ \\
	 & 
	 Identity testing \cite{valiant_automatic_2017,hangleiter2018sample} & TV distance & $1 \times n$ & classical & $O(\sqrt{2^n}/\epsilon^2)$& $O(\exp(n))$ & None \\
	 Shallow circuits & Local measurements \cite{hangleiter_direct_2017,NewSupremacy}& Fidelity witness&  $2 \times n$ & Hadamard and $T$ gate & $O(n^2/\epsilon^2)$ & $O(n^3/\epsilon^2)$ & Trusted single-qubit measurements \\\hline
     \multicolumn{7}{l}{\qquad \textbf{Quantum computation and simulation}}\\ \hline
      Fault-tolerant QC & Randomized benchmarking \cite{HarperEtAl2019} &Av.~gate~fidelity & $1 \times n$ & standard basis & $O(1/\epsilon^2_R)$ & $O(1/\epsilon^2_R)$ & i.i.d., SPAM robust \\
      & Gate set tomography \cite{blume-kohout_robust_2013} & Point estimates & $1\times n$ & standard basis & ? & ? & i.i.d., fully self-consistent \\
      State preparation &  Maximum-likelihood tomography~\cite{Hradil:1997tt} & Point estimate & $\exp(O(n)) \times n$ & ? & ? &  ?& i.i.d., trusted measurements\\
	  & Error-bar tomography~\cite{Wang:2019kp}& Confidence polytope & $\exp(O(n)) \times n$ & ? & ? & ?&i.i.d., trusted measurements\\
	  & Rank-$r$ tomography with $2$-designs \cite{2018arXiv180911162G}& Point estimate + confidence cert. & $\exp(O(n)) \times n$ & ? & $\tilde{O}(2^n r^2 / \epsilon^2)$ & $O(2^{3n})$ &i.i.d., trusted measurements\\
      & Direct fidelity estimation \cite{FidelityEstimation} & Fidelity estimate &$O(1/\epsilon^2) \times n$ & Pauli obs.& $O(2^n/\epsilon^2) $ &$O(2^n/\epsilon^2)$&i.i.d., trusted measurements\\
      & Cross-platform verification \cite{elben_cross-platform_2019} &Fidelity est.\ on $n_a$-qubit subsys. &  $ O(e^{b n_a}) \times n_a$, \quad  $b \lesssim 1$ & arb.~$1$-qubit unitaries & $O(e^{b n_a})$ & $ O (e^{b n_a})$ & i.i.d., trusted measurements\\
      &Self-testing $n$ Bell pairs \cite{natarajan_low-degree_2018}  & Fidelity witness & $2 \times n $ & $2$ conjugate bases & $1$  &$O(n)$ &Non-communicating and spatially separated observers \\
      Dynamical quantum simulation & MPO tomography \cite{baumgratz_scalable_2013,holzapfel_scalable_2015} & Point estimate + fidelity witness & $O(n) \times n$ & local operator basis & $O(n^3)$ & $\operatorname{poly}(n)$ &i.i.d., trusted measurements, MPO description\\
      & Building trust \cite{trotzky_probing_2012,ValidatingSimulator,schreiber_observation_2015,braun_emergence_2015} & Trust &?  & ? & ? & ? & Plenty\\
      \toprule
	\end{tabular}
	\caption{
	\label{thetable}
	We assess exemplary certification, characterization and benchmarking protocols for selected applications according to our framework: 
	the information or certificate that can be obtained from the protocol, the effort or complexity divided into its different parts, and the assumptions made. 
	We do so with respect to the number of qubits or optical modes, $n$, the depth of a circuit, $D$, and  additive $\epsilon$ or relative $\epsilon_R$ error tolerances.
	We write the measurement complexity as the number of distinct measurements $\times$ the number of single qubits or optical modes those measurements act on. 
	The sample complexity is the total number of samples required for the respective benchmark or certificate, we denote the Shannon entropy by $H$ and abbreviate `total-variation distance' by `TV distance'.
	We refer to \textcolor{orange}{BOX\,4} for guidance to the table.
	Where to the best of our knowledge no explicit resource scaling is in the literature we have put ?-marks. 
	}
\end{table*}
\end{turnpage}
\vspace*{-.15cm}
\newpage
\bibliographystyle{plain}

\end{document}